\documentclass[oneside]{article}

\usepackage{graphicx}
\usepackage[T2A]{fontenc}
\usepackage[cp1251]{inputenc}
\usepackage{cmpj2e,amsmath,amssymb}

\newcounter{Com}
\newcounter{Task}

\title
{Quasiparticles in the $XXZ$ model}

\author[P. Lu, G. M{\"u}ller, M. Karbach]{Ping Lu\refaddr{label1},
        Gerhard M{\"u}ller\refaddr{label1}, and Michael Karbach\refaddr{label1,label2}}

\addresses{
\addr{label1} Department of Physics, University of Rhode Island, Kingston RI
02881, USA
\addr{label2} Fachbereich Physik, Bergische Universit{\"a}t Wuppertal, 42097
Wuppertal, Germany
}

\begin{document}

\maketitle

\begin{abstract}
  The coordinate Bethe ansatz solutions of the $XXZ$ model for a
  one-dimensional spin-1/2 chain are analyzed with focus on the statistical
  properties of the constituent quasiparticles. Emphasis is given to the
  special cases known as $XX$, $XXX$, and Ising models, where considerable
  simplifications occur. The $XXZ$ spectrum can be generated from separate
  pseudovacua as configurations of sets of quasiparticles with different
  exclusion statistics. These sets are complementary in the sense that the
  pseudovacuum of one set contains the maximum number of particles from the
  other set. The Bethe ansatz string solutions of the $XXX$ model evolve
  differently in the planar and axial regimes. In the Ising limit they become
  ferromagnetic domains with integer-valued exclusion statistics. In the $XX$
  limit they brake apart into hard-core bosons with (effectively) fermionic
  statistics. Two sets of quasiparticles with spin 1/2 and fractional
  statistics are distinguished, where one set (spinons) generates the $XXZ$
  spectrum from the unique, critical ground state realized in the planar
  regime, and the other set (solitons) generates the same spectrum from the
  twofold, antiferromagnetically ordered ground state realized in the axial
  regime. In the Ising limit, the solitons become antiferromagnetic domain
  walls. 
\pacs 75.10.-b
\end{abstract}

\section{Introduction}\label{sec:intro}
Quantum spin chains are physically realized in quasi-one-dimensional magnetic
compounds.  These are crystalline materials with magnetic ions arranged in
exchange-coupled chains that are isolated from each other by non-magnetic ions.
The elementary magnetic moments are localized on the sites $l$ of a
one-dimensional lattice, which makes them distinguishable.  The associated spin
operators thus commute if they belong to different sites of that lattice,
$[S_{l}^{\alpha},S_{l'}^{\beta}]=\imath\hbar \epsilon_{\alpha\beta\gamma}
S_{l}^{\gamma} \delta_{ll'}$. The Hilbert space of a spin-1/2 chain with $N$
sites is conveniently spanned by product basis vectors
$|\sigma_{1}\dots\sigma_{N}\rangle$, ${\sigma_{l}=\uparrow, \downarrow}$.

Prominent among the many models employed in the context of quantum spin chain
compounds is the spin-1/2 $XXZ$ model,
\begin{equation}
  \label{eq:15plj3}
  \mathcal{H}_{XXZ} = J \sum_{l=1}^N \left\{ {S}^x_l {S}^x_{l+1} + 
    {S}^y_l {S}^y_{l+1} + \Delta {S}^z_l {S}^z_{l+1} \right\}.   
\end{equation}
It describes a uniform nearest-neighbor exchange coupling with uniaxial anisotroy.
Periodic boundary conditions are assumed. We distinguish ferromagnetic coupling
$(J<0)$ from antiferromagnetic coupling $(J>0)$, and the planar regime
$(0\leq\Delta<1)$ from the axial regime $(\Delta>1)$. Important special cases
are the $XX$ model $(\Delta=0)$, the $XXX$ model $(\Delta=1)$, and the Ising
model $(\Delta\to\infty)$.

One persistent challenge through decades of experiments on quantum spin chain
compounds has been the interpretation of the observed intensity spectrum in
terms of constituent quasiparticles \cite{SVW76,Furr00}.  The chief motivation
of the work reported here is to shed new light on this very issue. The approach
taken is eclectic in nature, combining older and more recent results for the
unified purpose of understanding the quasiparticle composition of the $XXZ$
spectrum more thoroughly.

The strong interest in the $XXZ$ model is sustained not only by its direct
relevance in experimental physics, but also by its amenability to exact
analysis via Bethe ansatz. The solution of the $XXX$ model was, in fact, the
very problem for which Bethe originally invented the method in the early days
of quantum mechanics \cite{Beth31,Batc07}. The Bethe ansatz allows for a
characterization of all many-body eigenstates as composed of quasiparticles
that scatter off each other or form bound states and thus turn into different
quasiparticles.  Over the years the same basic idea has been successfully
applied to many different kinds of systems and situations including interacting
boson and fermion gases with contact interactions \cite{LL63,YY69,Taka99}, the
Hubbard model for electrons on a lattice with on-site repulsion
\cite{LW68,EFG+05}, and for bosonic and fermionic quantum field theories
including the quantum Sine-Gordon and Thirring models \cite{BT79}.

\begin{figure}[htb]
\centerline{\includegraphics[width=12cm,angle=0]{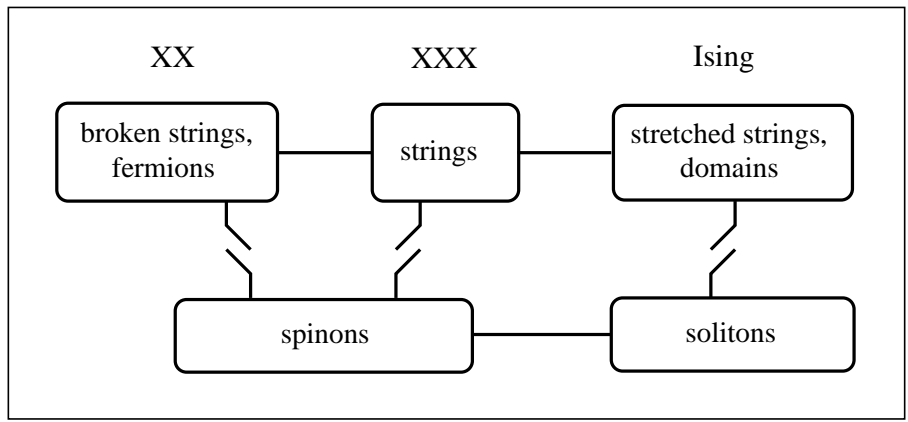}}
\caption{Zoo of quasiparticles that play some role in the context of the $XXZ$
  model.  For $J>0$, the pseudovacua of spinons and solitons are at the bottom
  of the spectrum. The pseudovacuum of the string particles is at the top in
  the axial regime ($\Delta \geq 1$) and moves downward in the planar regime
  ($0\leq\Delta\leq 1$), reaching the center in the $XX$ limit and coinciding there with
  the pseudovacuum of Jordan-Wigner fermions.}
\label{fig:qpzoo}
\end{figure}

The $XXZ$ spectrum can be generated from different pseudovacua by the
systematic creation of quasiparticles with different exclusion statistics.  In
most cases these pseudovacua are states of lowest energy (physical vacua). An
overview of the different kinds of quasiparticles that have emerged from
analytic work on the $XXZ$ model is shown in Fig.~\ref{fig:qpzoo}.  The boxes
in the top row represent quasiparticles whose names are derived from jargon
used to describe coordinate Bethe ansatz solutions. These quasiparticles are
characterized by strings of complex momenta with common real part and different
imaginary parts \cite{Taka99}. One-strings are unbound magnons, two-strings are
bound magnon pairs etc. Small (large) imaginary parts indicate loose (tight)
binding \cite{KM97}. At $\Delta=1$ the strings are constituent particles of
(degenerate) multiplets of eigenstates with total spin $S_T$. These multiplets
split up energetically at $\Delta \neq1$.

At $\Delta=0$ the energy levels join up in new degenerate configurations,
reflecting different symmetries \cite{BKMW04,AKMW06,KMW08,DFM01,FM01,FM01a}.  A
more natural classification scheme for the $XX$ spectrum is then based on
string fragments.  Hence the name {\it broken strings}.  These fragments are
closely related to the free lattice fermions that emerge from the Jordan-Wigner
representation of spin-1/2 operators \cite{BKMW04,DFM01,FM01,FM01a}. At
$\Delta>1$ the imaginary parts of the string solutions grow in magnitude and
diverge as $\Delta\to\infty$. Hence the name {\it stretched strings}. The
tightly bound strings are related, in the Ising limit, to localized domains of
reversed spins on successive lattice sites. The exclusion statistics of these
domains is similar to yet subtly different from that of the strings.

The boxes in the bottom row of Fig.~\ref{fig:qpzoo} represent quasiparticles
that are complementary to the string particles. The string pseudovacuum
contains the maximum number of spinons or solitons. The spinon and soliton
vacua contain strings at maximum capacity. Whereas the string particles have
integer-valued exclusion statistics, the spinons and solitons are realizations
of fractional statistics \cite{Hald91a}. They are both semions but with different
pseudovacua. The exclusion principle for semions is, roughly speaking, halfway
between those applicable for fermions and bosons. If it takes $1/g$ particles
to lower the number of orbitals in a band available for occupancy by one, then
$g=1$ describes fermions, $g=1/2$ semions, and $g=0$ bosons (as a limit).

The string particles and the semionic particles are natural building blocks for
the systematic construction of a complete $XXZ$ eigenbasis from different
pseudovacua. The configurations of string particles and semionic particles are
both unique and preserved in every $XXZ$ eigenstate. The ground state (physical
vacuum) of the $XXZ$ antiferromagnet $(J>0)$ coincides with the pseudovacua of
the semions. The spinon vacuum is unique and coincides with the non-degenerate
ground state of the $XXZ$ model at $0\leq \Delta\leq1$ for even $N$. The
soliton vacuum is twofold and coincides with the ground state of the $XXZ$
model at $\Delta>1$ for $N\to\infty$.

\begin{figure}[htb]
\centerline{\includegraphics[width=12cm,angle=0]{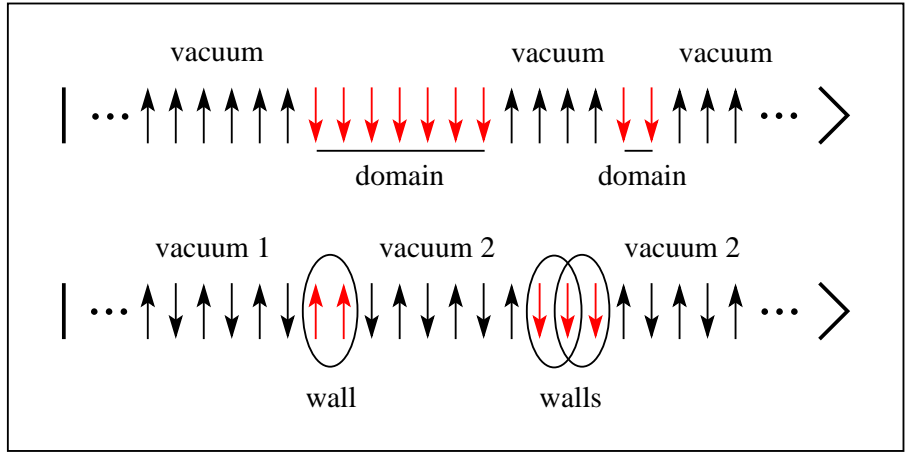}}
\caption{Constituent quasiparticles of $\mathcal{H}_{XXZ}$ at
  $\Delta\to\infty$ in real space (lattice of $N$ sites). They are either
  domains of flipped spins embedded in the unique vacuum with all spins up
  (top) or domain walls embedded in the twofold vacuum with all spins
  alternatingly up/down (bottom). }
\label{fig:qpis22ar}
\end{figure}

Let us briefly illustrate the relation between the particles from the top and
bottom rows in Fig.~\ref{fig:qpzoo} with two simple scenarios, one in
configuration space and the other in momentum space.  In the Ising limit,
$\mathcal{H}_{XXZ}$ has simple product eigenstates. The ferromagnetic state
$|\uparrow\uparrow\uparrow\cdots\rangle$ is the unique pseudovacuum for domains
of consecutive flipped spins, $\downarrow\downarrow\cdots\downarrow$, as shown
in top part of Fig.~\ref{fig:qpis22ar}. The twofold N{\'e}el state,
$|\uparrow\downarrow\uparrow\cdots\rangle$,
$|\downarrow\uparrow\downarrow\cdots\rangle$, by contrast, is the pseudovacuum
for antiferromagnetic domain walls of the kind $\uparrow\uparrow$,
$\downarrow\downarrow$ as shown in the bottom part of Fig.~\ref{fig:qpis22ar}.
These domain walls (named solitons) have effective spin $\pm1/2$.  The
integer-valued exclusion statistics of domains is associated with the fact that
neighboring particles must be separated by any positive integer number of
lattice sites. The fractional exclusion statistics of domain walls, on the
other hand, is associated with the fact that one of two lattice sites may be
shared by neighboring particles.

\begin{figure}[htb]
\centerline{\includegraphics[width=12cm,angle=0]{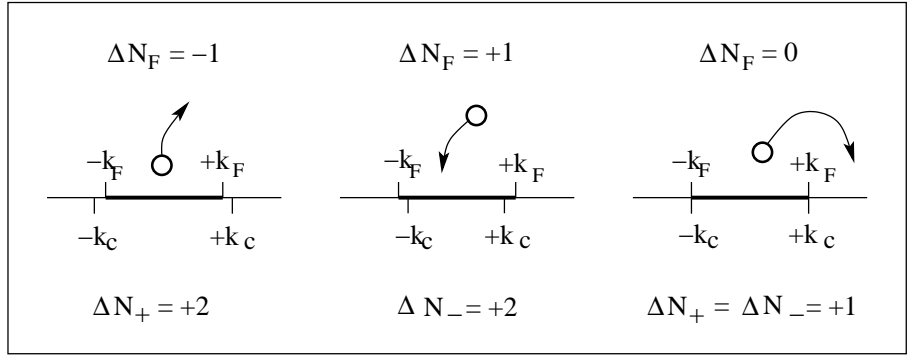}}
\caption{Constituent quasiparticles of $\mathcal{H}_{XXZ}$ at $\Delta=0$ in
  reciprocal space (fermionic band with Fermi momentum $k_{F}$). They are
  either spinless Jordan-Wigner fermion or semionic spinons with spin $\pm1/2$.}
\label{fig:xxfermvspin}
\end{figure}

In the $XX$ limit, $\mathcal{H}_{XXZ}$ is equivalent to a system of $N_{F}$
free lattice fermions in a band \cite{LSM61,Kats62,MBA71,Derz02}. The ground state
corresponds to the intact Fermi sea at $|k|<k_{F}$ as shown in
Fig.~\ref{fig:xxfermvspin}. The entire spectrum can be generated systematically
via particle excitations $(\Delta N_{F}=+1)$, hole excitations $(\Delta
N_{F}=-1)$, and particle-hole excitations $(\Delta N_{F}=0)$.  The Fermi-sea
ground state can be viewed as the pseudovacuum for semionic spinons. We
introduce a threshold momentum $k_{c}$ that varies with $N_{F}$. When a fermion
is removed from the intact sea, $k_{c}$ slightly increases to generate not just
one but two vacancies in the interval $|k|<k_{c}$ of the band. These two holes
are then identified with a pair of spin-up spinons.  When a fermion is added
outside the intact Fermi sea, $k_{c}$ slightly decreases to produce two
particles in the region $|k|>k_{c}$. They are identified with a pair of
spin-down spinons. A fermionic particle-hole excitation leaves $k_{c}$
unchanged and is interpreted as a pair of spinons with opposite spin
orientation. The entire spectrum can thus be described in the form of spinon
configurations.

The main goal of this work is to identify the relationship between the string
particles and the semionic particles for $\mathcal{H}_{XXZ}$ at
$\Delta=0,1,\infty$ in particular and to illuminate how this relationship can
be tracked between the three points in parameter space. In Sec.~\ref{sec:BA}
we set the stage for this investigation by reviewing the relevant Bethe-ansatz
representations that best serve our purpose. We then discuss the quasiparticle
composition of the $XXZ$ spectrum in the $XXX$ limit (Sec.~\ref{sec:XXX}), the
Ising limit (Sec.~\ref{sec:Ising}), and the $XX$ limit (Sec.~\ref{sec:XX}).

\section{Bethe ansatz}\label{sec:BA}
The coordinate Bethe ansatz for $\mathcal{H}_{XXZ}$ uses the magnon vacuum
$|F\rangle\equiv|\uparrow\uparrow\cdots\uparrow\rangle$ with energy $E_F=
JN\Delta/4$ as its reference state. The Bethe form \cite{Beth31} of any
eigenvector in the invariant subspace with magnetization $M_{z}=N/2-r$,
\begin{equation}\label{eq:psirplj3}
 |\psi\rangle = \sum_{1\leq n_1<\ldots<n_r\leq N} a(n_1,\ldots,n_r)
 S_{n_1}^-\cdots S_{n_r}^-|F\rangle,
\end{equation}
has coefficients of the form
\begin{equation}\label{eq:barplj3}
 a(n_1,\ldots,n_r) =
 \!\sum_{{\cal P}\in S_r}
 \!\exp\left(\!\imath\sum_{j=1}^r k_{{\cal P} j}n_j
 + \frac{\imath}{2}\sum_{i<j}^{r} \theta_{{\cal P}i{\cal P}j}\!\right),
\end{equation}
determined by $r$ magnon momenta $k_i$ and one phase angle,
$\theta_{ij}=-\theta_{ji}$, for each magnon pair. The sum ${\cal P}\in S_r$ is
over the permutations of the indices $\{1,2,\ldots,r\}$.
The $k_i$ and $\theta_{ij}$ satisfy the Bethe ansatz equations (BAE),
\begin{equation}
  \label{eq:16plj3}
  e^{\imath N k_i} =  \prod_{j \neq i}^re^{\imath\theta_{ij}}, \qquad
 e^{\imath\theta_{ij}}= - \frac{e^{\imath(k_i + k_j)} + 1 - 2 \Delta e^{\imath k_i}}%
  {e^{\imath (k_i + k_j)} + 1 - 2 \Delta e^{\imath k_j}}.
\end{equation}
The energy and the wave number of an eigenstate with magnon momenta
$\{k_{1},\ldots,k_{r}\}$ are
\begin{equation}
  \label{eq:17plj3}
  \frac{E-E_F}{J} =  \sum_{i=1}^r \left( \cos k_i - \Delta \right),\qquad 
  k = \left(\sum_{i=1}^r k_i\right) \mathrm{mod}(2\pi).
\end{equation}

Reasons of practicality dictate the use of different transformations of the BAE
in the regimes of axial anisotropy $(\Delta>1)$, planar anisotropy
$(\Delta<1)$, and isotropic exchange $(\Delta=1)$.  It is convenient to
introduce rapidities $z_{i}$,
\begin{align}
  \label{eq:6}
  \cot\frac{k_{i}}{2} =
  \begin{cases} \displaystyle
    \cot\frac{\gamma}{2} \tanh\frac{\gamma z_{i}}{2} 
    &:  0 \leq \Delta = \cos\gamma < 1 
    \\ \displaystyle
    z_{i}
    &: \Delta = 1
    \\ \displaystyle
    \coth\frac{\eta}{2} \tan\frac{\eta z_{i}}{2} 
    &:  \Delta = \cosh\eta >1 
  \end{cases},
\end{align}
which renders the limit $\Delta\to1$ smooth from both sides. The BAE
\eqref{eq:16plj3} thus transform into
\begin{align}
  \label{eq:15}
  N \tilde{\phi}_{1}(z_i) = 
  2 \pi I_i + \sum_{j \neq i}^r \tilde{\phi}_{2}(z_i-z_j),  
    \quad i = 1,\ldots,r, 
\end{align}
where
\begin{align}
  \label{eq:7}
  \tilde{\phi}_{\nu}(z) \doteq 
  \begin{cases}\displaystyle
  2\arctan\left(\cot\frac{\gamma\nu}{2}\tanh\frac{\gamma z}{2}\right)
  &:  0\leq\Delta  < 1 
  \\ \displaystyle
  2\arctan\frac{z}{\nu}
  &:  \Delta = 1 
  \\ \displaystyle
  2\arctan\left(\coth\frac{\eta\nu}{2}\tan\frac{\eta z}{2} \right)
  &:  \Delta  > 1 
  \end{cases}.
\end{align}
The Bethe quantum numbers (BQN) $I_{i}$ of integer or half-integer value
reflect the multivaluedness of the logarithm used in the transformation. They
are subject to restrictions that will be discussed case by case. The energy
expression in \eqref{eq:17plj3} becomes
\begin{align}
  \label{eq:16}
  \frac{E-E_F}{J} = -\sum_{i=1}^r \tilde{e}(z_{i}),\qquad \tilde{e}(z)\doteq
 \begin{cases}\displaystyle
     \frac{\sin^{2}\gamma}{\cosh(\gamma z)-\cos\gamma}&:
    0\leq\Delta <  1 \\ \displaystyle
    \frac{2}{1+z^{2}} &: \Delta = 1 \\\displaystyle
    \frac{-\sinh^{2}\eta}{\cos(\eta z)-\cosh\eta} &:
    \Delta>1 
  \end{cases}.
\end{align}

Note that we are, effectively, dealing with a single anisotropy parameter,
$\eta=\imath\gamma$, that is real in one regime, imaginary in the other, and
zero at the isotropy point. This parametrization is particularly useful for
tracking the spectrum between the axial and planar regimes across the point of
higher rotational symmetry. Slightly different parametrizations are more
adequate for the exploration of the limits $\Delta\to\infty$ in the axial regime
and $\Delta\to0$ in the planar regime.

\subsection{Axial regime}\label{sec:axial}
At $\Delta>1$ we use the transformation \cite{CG66},
\begin{equation}
  \label{eq:18plj3}
  \tan \frac{z_i}{2} \doteq \tanh \frac{\eta}{2} \cot \frac{k_i}{2}, 
  \qquad
  \eta = \mathrm{arcosh}\,\Delta,
  \qquad
  -\pi < z_{i} < \pi,
\end{equation}
instead of (\ref{eq:6}), to bring the BAE (\ref{eq:16plj3}) into the form 
\begin{equation}
  \label{eq:19plj3}
  \left( \frac{\coth(\eta/2) \tan(z_i/2) - \imath}
              {\coth(\eta/2)\tan(z_i/2) + \imath} \right)^N 
  = \prod_{j \neq i}^r 
  \frac{\coth(\eta) \tan \left[(z_i - z_j)/2\right] - \imath}%
       {\coth(\eta) \tan \left[(z_i - z_j)/2\right] + \imath}, 
       \qquad i = 1,\dots,r.
\end{equation}
The associated trigonometric BAE are
\begin{equation}
  \label{eq:20plj3}
  N \phi_{1}(z_i) = 
  2 \pi I_i + \sum_{j \neq i}^r \phi_{2}(z_i-z_j),  
    \quad i = 1,\ldots,r, 
\end{equation}
where 
\begin{align}
  \label{eq:13}
  \phi_{\nu}(z) \doteq 2\arctan
  \left(\frac{\tan(z/2)}{\tanh(\eta\nu/2)}\right)
  + 2\pi
  \left\lfloor
    \frac{\Re z}{2\pi}+\frac{1}{2}
  \right\rfloor.
\end{align}
The second term in (\ref{eq:13}) ensures that the set $\{I_{i}\}$
remains the same as an eigenstate is tracked across the axial regime
\cite{Taka99}. Here $\lfloor x\rfloor$ is the floor function (integer part of
$x$). 

For the analysis of solutions that include complex magnon momenta we invoke
the string hypothesis for the rapidities \cite{Taka99}:
\begin{equation}
  \label{eq:104plj3}
  z_{\alpha}^{m,l} =z^{m}_{\alpha} + \imath\eta(m+1-2l),\qquad l=1,\ldots,m,
  \qquad m=1,2,\ldots,r.
\end{equation}
The index $l$ distinguishes rapidities belonging to the same string (of size
$m$). The index $\alpha =1,...,n_{m}$ distinguishes different strings of the
same size.  The string ansatz (\ref{eq:104plj3}) produces only asymptotic
solutions of the BAE (\ref{eq:19plj3}) for $N\to\infty$. The finite-$N$
corrections are, in general, exponentially small, and not all finite-$N$
solutions fit the string template (\ref{eq:104plj3}) \cite{Taka99}. However,
neither corrections nor exceptions affect macroscopic systems. In the Ising
limit, where the spread of imaginary parts in (\ref{eq:104plj3}) diverges, all
corrections and exceptions disappear even for finite $N$.

A given string solution of (\ref{eq:19plj3}) with magnetization $M_{z}=N/2-r$
is described by $r$ rapidities that breaks down into configurations of strings
such that the constraint,
\begin{equation}
  \label{eq:105plj3}
  \sum_{m\in\mathcal{C}} m\, n_{m}=r,
\end{equation}
is satisfied, where the set $\mathcal{C}$ identifies those sizes of strings
that occur in a given eigenstate. With the functions
\begin{equation}
  \label{eq:71mk}
  \varphi_{\nu}(z) 
   \doteq 
  \frac{\coth(\eta\nu/2)\tan(z/2)-\imath}
       {\coth(\eta\nu/2)\tan(z/2)+\imath}
  =\frac{\sin\Big((z-\imath\eta\nu)/2\Big)}
        {\sin\Big((z+\imath\eta\nu)/2\Big)}
\end{equation}
we rewrite Eqs.~(\ref{eq:19plj3}) in the form
\begin{equation}
  \label{eq:106plj3}
  \left[\varphi_{1}(z_{i})\right]^{N} 
  = \prod_{j \neq i}^r \varphi_{2}(z_{i}-z_{j}), 
  \quad i = 1,\dots,r.
\end{equation}
When we substitute the string ansatz (\ref{eq:104plj3}) into Eqs.~(\ref{eq:106plj3})
we obtain
\begin{equation}
  \label{eq:107plj3}
  \left[\varphi_{1}(z_{\alpha}^{m,l})\right]^{N} 
   = \prod_{(m',\beta) \neq (m,\alpha)}\prod_{k=1}^{m'}\;
  \varphi_{2}(z_{\alpha}^{m,l}-z_{\beta}^{m',k}) 
  \prod_{k\neq l}^{m}
  \varphi_{2}(z_{\alpha}^{m,l}-z_{\alpha}^{m,k}),
\end{equation}
for $l =1,...,m$ and $\alpha=1,\ldots,n_{m}$, where
\begin{equation}
  \label{eq:108plj3}
  \varphi_{\nu}(z_{\alpha}^{m,l})
  =\frac{\sin\Big([z_{\alpha}^{m}+\imath\eta(m+1-\nu-2l)]/2\Big)}%
        {\sin\Big([z_{\alpha}^{m}+\imath\eta(m+1+\nu-2l)]/2\Big)}.
\end{equation}
To determine the real parts, $z_{\alpha}^{m}$, we form the
product of all Eqs.  (\ref{eq:107plj3}) for fixed $m,\alpha$:
\begin{equation}
  \label{eq:109plj3}
  \left[\prod_{l=1}^{m} \varphi_{1}(z_{\alpha}^{m,l})\right]^{N} 
  =\prod_{(m',\alpha') \neq (m,\alpha)}
  \left[\prod_{k=1}^{m'}\prod_{l=1}^{m}
  \varphi_{2}(z_{\alpha}^{m,l}-z_{\alpha'}^{m',k})\right] 
  \left[\prod_{l=1}^{m}\prod_{k\neq l}^{m}
  \varphi_{2}(z_{\alpha}^{m,l}-z_{\alpha}^{m,k})\right].
\end{equation}
Each expression in square brackets can be simplified massively, producing the BAE for
the $z_{\alpha}^{m}$,
 \begin{align}
   \label{eq:113plj3}
   \left[\varphi_{n}(z_{\alpha}^{m})\right]^{N} = 
   \prod_{(m',\alpha') \neq (m,\alpha)}
   \varphi_{m'-m}(z_{\alpha\alpha'}^{mm'}) \varphi_{m'+m}(z_{\alpha\alpha'}^{mm'})
   \prod_{l=1}^{m-1} \left[\varphi_{m'+m-2l}(z_{\alpha\alpha}^{mm'})\right]^{2},
 \end{align}
with $z_{\alpha\alpha'}^{mm'}\doteq z_{\alpha}^{m}-z_{\alpha'}^{m'}$.
The associated trigonometric BAE, 
\begin{equation}
  \label{eq:1}
  N\phi_{m}(z_{\alpha}^{m}) = 2\pi I_{\alpha}^{m}+
  \sum_{(m',\alpha') \neq (m,\alpha)} \Phi_{mm'}(z_{\alpha\alpha'}^{mm'}),
\end{equation}
\begin{align}
  \label{eq:8}
  \Phi_{mm'}(z) \doteq
  \begin{cases}
    \phi_{|m-m'|}(z)+2\phi_{|m-m'|+2}(z)+...+2\phi_{m+m'-2}(z)+\phi_{m+m'}(z)
    &: m' \neq m 
    \\
    2\phi_{2}(z)+2\phi_{4}(z)+...+2\phi_{2m-2}(z)+\phi_{2m}(z)
    &: m' = m 
  \end{cases},
\end{align}
depend on a set $\{I_{\alpha}^{m}\}$ of BQN that reflects the specific string
combination of any given eigenstate.
The energy and wave number of that state are 
\begin{align}
  \label{eq:18}
  \frac{E-E_{F}}{J} = -\sum_{(m,\alpha)} \frac{\sinh\eta\sinh(\eta
    m)}{\cosh(\eta m)-\cos z_{\alpha}^{m}},\qquad
  k=\left[\sum_{(m,\alpha)}\left(\pi-\frac{2\pi}{N}I_{\alpha}^{m}\right)
  \right] \mathrm{mod}(2\pi).
\end{align}
The range of the $I_{\alpha}^{m}$ will be discussed first for the case $\Delta=1$ in
Sec.~\ref{sec:isotropic} and then for the axial regime including the Ising
limit in Sec.~\ref{sec:Ising}.

\subsection{Isotropic exchange}\label{sec:isotropic}
At $\Delta=1$ we retain the rapidities from (\ref{eq:6}), the trigonometric
BAE in the form (\ref{eq:15}), and the energy expression (\ref{eq:16}).
The string hypothesis now reads
\begin{equation}
  \label{eq:5}
  z_{\alpha}^{m,l} =z^{m}_{\alpha} + \imath(m+1-2l),\qquad
  l=1,\ldots,m,\qquad m=1,2,\ldots,r.
\end{equation}
The BAE for the real parts $z_{\alpha}^{m}$ are Eqs.~\eqref{eq:113plj3}
with (\ref{eq:108plj3}) replaced by
\begin{equation}
  \label{eq:14}
  \varphi_{\nu}(z_{\alpha}^{m,l})
  =\frac{z_{\alpha}^{m}+\imath(m+1-\nu-2l)}%
        {z_{\alpha}^{m}+\imath(m+1+\nu-2l)}.
\end{equation}
The associated trigonometric BAE for the $z_{\alpha}^{m}$ then take on the form
\eqref{eq:1} with $\Phi_{mm'}(z)$ from (\ref{eq:8}) and $\phi_{\nu}(z)\doteq
2\arctan(z/\nu)$ from (\ref{eq:7}). The energy and wave number of
an eigenstate specified by the set $\{I_{\alpha}^{m}\}$ are
\begin{equation}
  \label{eq:9}
  \frac{E-E_{F}}{J} = -\sum_{(m,\alpha)} \frac{2m}{m^{2}+(z_{\alpha}^{m})^{2}},\qquad 
 k=\left[\sum_{(m,\alpha)}\left(\pi-\frac{2\pi}{N}I_{\alpha}^{m}\right)\right]
 \mathrm{mod}(2\pi). 
\end{equation}
The string hypothesis sets the range of the $I_{\alpha}^{m}$ (with
$I_{\alpha+1}^{m}>I_{\alpha}^{m}$ implied) as follows \cite{Taka99}:
\begin{equation}
  \label{eq:10}
  |I_{\alpha}^{m}|\leq \frac{1}{2}\left(N-1-\sum_{m'\in\mathcal{C}}t_{mm'}
n_{m'}\right),\qquad t_{mm'}\doteq 2\mathrm{min}(m,m') -\delta_{mm'},
\end{equation}
where $n_{m}$ is the number of $m$-strings (distinguished by running index
$\alpha$) in the eigenstate. Note that the range of allowed values becomes
narrower for all sizes if a string of any size is added. The $I_{\alpha}^{m}$
for a given combination $\{n_{m}\}$ and a given value of $m$ are either all
integers or all half-integers such that the border values of the range
(\ref{eq:10}) are realized. 

The Bethe state characterized by a set $\{I_{\alpha}^{m}\}$ is the
highest-weight component (i.e. the state with $M_{z}=S_{T}$) of an
$S_{T}$-multiplet with total spin
\begin{equation}
  \label{eq:11}
  S_{T}= \frac{N}{2}-r,\qquad r=\sum_{m\in \mathcal{C}}mn_{m}.
\end{equation}
The other components of any given $S_{T}$-multiplet are generated by the
addition of magnons with zero momentum. These have no effect on the energy or the
wave number. Consider the case of a highest-weight state with total spin
(\ref{eq:11}) that only contains 1-strings. Suppose this state is specified by
the following set of BQN subject to the constraint (\ref{eq:10}):
\begin{equation}
  \label{eq:28}
  -\frac{1}{2}(N-r-1)\leq I_{1}^{1}<\cdots<I_{r}^{1}\leq \frac{1}{2}(N-r-1).
\end{equation}
The rapidities $z_{\alpha}^{1}$ derived from the BAE (\ref{eq:1}),
\begin{equation}
  \label{eq:29}
  N\phi_{1}(z_{\alpha}^{1})= 2\pi I_{\alpha}^{1}
  +\sum_{\beta\neq\alpha}^{r}\phi_{2}(z_{\alpha}^{1}-z_{\beta}^{1}),\quad
  \alpha=1,\ldots,r, 
\end{equation}
are real and, with rare exceptions, finite. The member state with
$M_{z}=S_{T}-1$ of the same multiplet has one extra rapidity,
$z_{r+1}=\infty$, representing the additional magnon with $k_{r+1}=0$. The BAE
(\ref{eq:1}) for this state are then satisfied with the same $z_{\alpha}^{1}$,
$\alpha=1,\ldots,r$ and with $z_{r+1}=\pm\infty$ if we set the BQN as follows:
\begin{equation}
  \label{eq:30}
  \tilde{I}_{\alpha}^{1}=I_{\alpha}^{1}\pm \frac{1}{2},\quad
  \alpha=1,\ldots,r;\qquad \tilde{I}_{r+1}^{1}=\pm \frac{1}{2}(N-r).
\end{equation}
In the more general case, where the highest-weight state under consideration
contains strings with $m>1$, the shifts in the already existing BQN,
$I_{\alpha}^{m}$, and the value of the new BQN, $I_{r+1}^{1}$, will be
different.  In Sec.~\ref{sec:XXX} we will treat the strings as interacting
particles and examine their exclusion statistics.

\subsection{Planar regime}\label{sec:planar}
The string ansatz at $\Delta<1$ \cite{Taka99} will not be used here. 
Simplifications and residual complications that occur in the limit $\Delta\to0$
can be seen in the raw form (\ref{eq:16plj3}) of the BAE. There are two
categories of solutions, both of which are ubiquitous. Regular and singular
solutions are distinguished by the absence or presence of pairs of critical
magnon momenta with $k_{i}+k_{j}=\pi$, which make both the numerator and the
denominator in (\ref{eq:16plj3}) vanish as $\Delta\to0$.

All regular solutions produce real magnon momenta from $e^{\imath
  Nk_{i}}=(-1)^{r-1}$, whereas singular solutions include critical magnon pairs
that are either both real or form a complex-conjugate pair
\cite{BKMW04,DFM01,FM01,FM01a}. The latter can be interpreted as fragments of
strings that exist throughout the planar regime. Since all critical pairs are
associated with a twofold degeneracy of eigenstates with equal
wave number, the singular features as imposed by the BAE in the limit
$\Delta\to0$ can be removed by unitary transformations. This erases, at
$\Delta=0$, all traces of the string nature, at $\Delta>0$, of complex
solutions.  The magnon momenta thus regularized are
\begin{equation}
  \label{eq:33}
  k_{\alpha} =\pi-\frac{2\pi}{N}I_{\alpha}^{1}.
\end{equation}
The associated BQN are integers for odd $r$ and half-integers for even $r$ with
range
\begin{equation}
  \label{eq:34}
  |I_{\alpha}^{1}-\tau_{r}|\leq \frac{1}{2}(N-1),\qquad
  \tau_{r}=\frac{1}{2}[1-(-1)^{r}]. 
\end{equation}
Their relation to the BQN (\ref{eq:10}) and (\ref{eq:30}) of the
$S_{T}$-multiplet states will be discussed in Sec.~\ref{sec:XX}. 

The regularized BAE solutions (\ref{eq:33}) describe hard-core bosons. The
phase shift is $\theta_{ij}=\pi$ for all two-particle interactions. These
hard-core bosons are equivalent to the Jordan-Wigner fermions
\cite{LSM61,Kats62,MBA71,Derz02} that have been instrumental in most studies of
the $XX$ model.

\section{Quasiparticle composition of XXZ
spectrum}\label{sec:QCS}
We are now ready to explore the relationship between the complementary
quasiparticle compositions of the $XXZ$ spectrum. We begin in
Sec.~\ref{sec:XXX} with string particles and the complementary spinon particles
for the $XXX$ case. In Sec.~\ref{sec:Ising} we then discuss the effects of
axial anisotropy on the strings and their relationship to ferromagnetic domains
in the Ising limit.  Complementary to these domains are the soliton particles
in the shape of antiferromagnetic domain walls.  In the planar regime the
strings evolve differently. What remains of them in the $XX$ limit are
fragments that act like hard-core bosons or, equivalently, free Jordan-Wigner
fermions.  Complementary to the latter are again the spinons as will be
discussed in Sec.~\ref{sec:XX}.

\subsection{XXX limit: strings and spinons}\label{sec:XXX}
The very structure of the coordinate Bethe ansatz suggests that the strings
(\ref{eq:5}) can be interpreted as quasiparticles. There exists a universal
energy-momentum relation as implied by (\ref{eq:9}). The particle interaction
is encoded in the set of momenta (or rapidities) dictated by the BAE and in the
phase shifts associated with elastic two-particle collisions.

The exclusion statistics of strings is determined by the rule (\ref{eq:10})
governing the range of BQN and by the relation (\ref{eq:11}) governing the
capacity for strings in a highest-weight state of given $S_{T}$. The total
number of $S_{T}$-multiplets with string content $\{n_{1},n_{2},\ldots\}$
becomes the solution of a standard combinatorial problem:
\begin{equation}
  \label{eq:19}
  W(\{n_m\}) = \prod_{m\in\mathcal{C}}
\left( \begin{tabular}{c} $d_m+n_m-1$ \\ $n_m$ \end{tabular}\right),\qquad
d_m =A_m -\sum_{m'\in \mathcal{C}}g_{mm'}(n_{m'}-\delta_{mm'}),
\end{equation}
where
\begin{equation}
  \label{eq:20}
  A_m=N+1-2m,\qquad g_{mm'}=2\min(m,m')
\end{equation}
are statistical capacity constants and statistical interaction coefficients,
respectively, that are specific to the string particles
\cite{AKMW06,Hald91a,Wu94,FS98,LVP+08}. Taking into account the
$(2S_{T}+1)$-fold degeneracy of each multiplet, this classification accounts
for the complete spectrum,
\begin{equation}
  \label{eq:21}
  \sum_{\{n_m\}}W(\{n_m\})(2S_T+1)=2^N,
\end{equation}
with the dependence of $S_{T}$ on $\{n_{m}\}$ given in (\ref{eq:11}).

\begin{figure}[ht!]
\centerline{\includegraphics[width=10cm,angle=0]{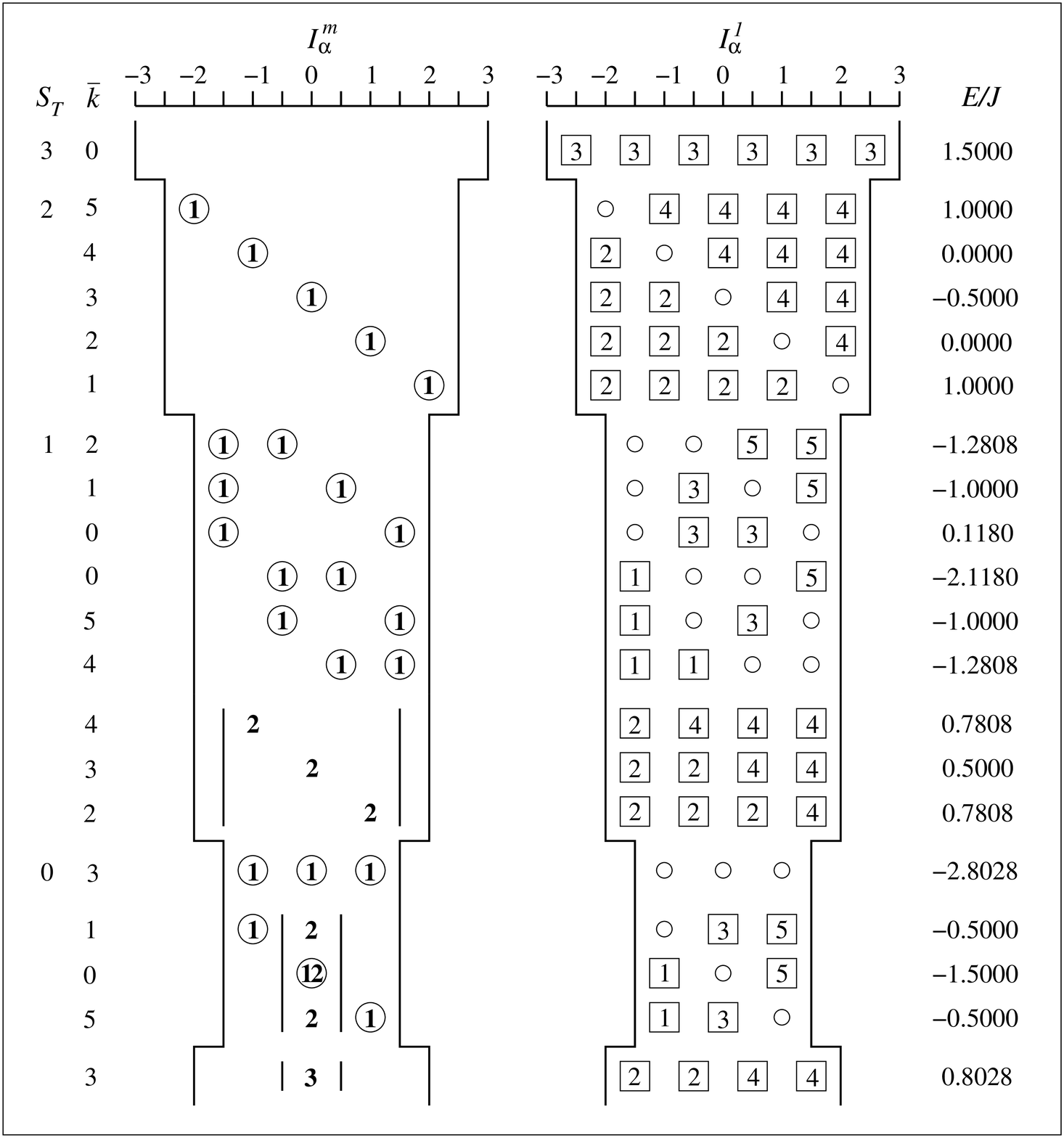}}
\caption{Specification of all $S_{T}$-multiplets of the $XXX$ model for $N=6$
  in the string representation (left) and spinon representation (right). The
  quantum numbers $S_{T}$ and $\bar{k}\doteq Nk/2\pi$ of each multiplet are
  stated on the far left. The positions of the numbers $m=1,2,3$ and the
  vertical lines (left) mark the values of the $I_{\alpha}^{m}$ and their
  range. The $I_{\alpha}^{1}$, emphasized by circles, are markers for the
  spinon configurations (right), where they are reproduced as small circles.
  Each square on the right represents a spinon. The set of squares is
  complementary to that of circles with the same range. The number
  $m_{i}=1,2,\ldots,5$ inside each square marks the spinon orbital. The energy
  of each $S_{T}$ multiplet is stated on the far right.}
\label{fig:stringN6}
\end{figure}

To illustrate the string composition of the $XXX$ spectrum and to explain its
relationship to the complementary spinon composition we consider a chain of
$N=6$ sites. In Table~\ref{tab:comstrN6} we list all combinations $\{n_{m}\}$
permitted by (\ref{eq:11}). Also listed is the range of all $I_{\alpha}^{m}$
present in each combination as inferred from (\ref{eq:10}). The distinct
configurations of $I_{\alpha}^{m}$ thus allowed produce the highest-weight
states of all $S_{T}$-multiplets as shown in Fig.~\ref{fig:stringN6} (left).
The quantum number $S_{T}$ depends on the combinations $\{n_{m}\}$ via
(\ref{eq:11}) and the wave number $k$ depends on the configurations
$\{I_{\alpha}^{m}\}$ via (\ref{eq:9}).  Each row of $I_{\alpha}^{m}$ in
Fig.~\ref{fig:stringN6} (left) represents the distinct string motif of an
$S_{T}$-multiplet. The solution of the BAE (\ref{eq:1}) thus specified is for
the highest-weight component of the multiplet. The non-highest-weight
components are characterized by additional BQN as described at the end of
Sec.~\ref{sec:isotropic}. The role of these additional BQN will be discussed in
Sec.~\ref{sec:Ising} for $\Delta>1$ (see Fig.~\ref{fig:stringN6plus}) and in
Sec.~\ref{sec:XX} for $\Delta<1$ (see Fig.~\ref{fig:stringN6minus}).

\begin{table}[htb]
  \caption{String composition $\{n_{m}\}$ of all
    $S_{T}$-multiplets for $N=6$. Each row describes a distinct combination for
    a total of 7. Each combination produces $W$ multiplets for a total of
    20. Each multiplet represents $2S_{T}+1$ states for a total of 64. Associated
    with each string in a given combination is a BQN $I_{\alpha}^{m}$ of the range
    shown.} 
\label{tab:comstrN6}  
\begin{center}\vspace*{2mm}
\begin{tabular}{c|ll|cc|l}
$r$ & $\mathcal{C}$ & $n_m$ & $W$ & $2S_{T}+1$ & BQN  \\ \hline
$0$ & -- & -- & $1$ & $7$ & --   \\
$1$ & $\{1\}$ & $1$ & $5$ & $5$ & $|I_\alpha^{1}|\leq2$  \\ 
$2$ & $\{1\} $& $2$ & $6$ & $3$ & $|I_\alpha^{1}|\leq\frac{3}{2}$  \\ 
$2$ & $\{2\}$ & $1$  & $3$ & $3$ & $|I_\alpha^{2}|\leq1$ \\ 
$3$ & $\{1\}$ & $3$ & $1$ & $1$ & $|I_\alpha^{1}|\leq1$ \\ 
$3$ & $\{1,2\}$ & 1,1 & $3$ & $1$ & $|I_\alpha^{1}|\leq1, |I_\alpha^{2}|=0 $ \\ 
$3$ & $\{3\}$ & $1$  & $1$ & $1$ & $|I_\alpha^{3}|=0$
\end{tabular}
\end{center}
\end{table} 

Elements of the string motif serve as the template of the spinon motif of the
same $S_{T}$-multiplet.  Spinons are specified by their spin and momentum
quantum numbers. The spinon interaction depends on the particle momenta and (in
general) also on the particle spins.  The unique $XXX$ ground state for even
$N$ is the spinon vacuum. The ground state for odd $N$ is fourfold degenerate
and contains exactly one spinon. The total number $N_{s}$ of spinon is
restricted to be even (odd) for even (odd) $N$ and can assume the values $0\leq
N_{s}\leq N$. Counting the spin-up spinons $(N_{+})$ and spin-down spinons
$(N_{-})$ separately produces the relations
\begin{equation}
  \label{eq:22}
  N_{+}+N_{-}=N_{s},\qquad N_{+}-N_{-}=2M_{z}.
\end{equation}

It is useful to introduce spinon orbitals associated with distinct spinon
momentum quantum numbers $m_{i}$.  The wave number of any $XXX$ multiplet can
be expressed in terms of the spinon orbital momenta as follows:
\begin{equation}
  \label{eq:24}
  k= \left(\frac{\pi}{N}\sum_{j=1}^{N_{s}}m_{j} -\frac{N\pi}{2}\right) \mathrm{mod}(2\pi).
\end{equation}
The range of $m_{i}$ depends on both $N$ and $N_{s}$:
\begin{equation}
  \label{eq:23}
m_{i}= \frac{N_{s}}{2}, \frac{N_{s}}{2}+2, \ldots, N-\frac{N_{s}}{2}.
\end{equation}
The number of available orbits to $N_{s}$ spinons,
$N_{\mathrm{orb}}=(N-N_{s})/2+1$, thus decreases by one for every two spinons
added, $\Delta N_{\mathrm{orb}}=-\Delta N_{s}/2$, which is one way of
recognizing the semionic nature of spinon particles. The exact exclusion statistics
of spinons is encoded in the number of ways $N_{+}$ spin-up spinons and
$N_{-}$ spin-down spinons can be distributed among $N_{\mathrm{orb}}$
accessible orbitals. The spinon multiplicity expression \cite{Hald91a},
\begin{equation}
  \label{eq:25}
  W(N_{+},N_{-}) = \prod_{\sigma=\pm}
  \left( \begin{tabular}{c} $d_\sigma+N_\sigma-1$ \\ $N_\sigma$
    \end{tabular}\right),\qquad 
d_\sigma = A_\sigma-\sum_{\sigma'=\pm}g_{\sigma\sigma'}(N_{\sigma'}-\delta_{\sigma\sigma'}),
\end{equation}
\begin{equation}
  \label{eq:26}
 A_\sigma=\frac{1}{2}(N+1),\qquad  g_{\sigma\sigma'}=\frac{1}{2},
\end{equation}
is cast in the same general form as its string counterpart (\ref{eq:19}).
Summation of $W(N_{+},N_{-})$ over all allowed values of $N_{\pm}$ accounts for
all $2^{N}$ states.

The complementary relationship between the string and spinon particles is
reflected in their motifs. In Fig.~\ref{fig:stringN6} we show the string motif
and spinon motif of all 20 $S_{T}$-multiplets for $N=6$ side by side.  The
number of spinons contained in each eigenstate of a given $S_{T}$-multiplet is
equal to the number of vacancies left by the 1-string BQN, $I_{\alpha}^{1}$,
across the range (\ref{eq:10}). In Fig.~\ref{fig:stringN6} (right) we have
marked the positions of the $I_{\alpha}^{1}$ by circles and the vacancies by
squares. The number inside each square denotes the orbital $m_{i}$ to which
every spinon belongs. The available orbitals depend on $N_{s}$ via
(\ref{eq:23}).

The rules for assigning spinons of particular momentum quantum numbers to
orbitals must ensure (i) that the $m_{i}$ reproduce, via (\ref{eq:24}), the
wave number $k$ already known via (\ref{eq:9}) and (ii) that the permissible
spin orientations are consistent with the quantum number $S_{T}$. The
allowed combinations of spinon orbitals are encoded in those
$S_{T}$-multiplets that do not contain any strings with $m>1$. Here the
positions of the 1-strings, i.e. the circles in the spinon motifs, play the
role of delimiters between successive spinon orbitals.

A spinon orbital with $l_{i}$ spinons has an orbital spin
$S_{i}^{\mathrm{orb}}=l_{i}/2$, $i=1,\ldots,N_{\mathrm{orb}}$. If all $N_{s}$
spinons are in the same orbital such as in the first two motifs (numbered from
top to bottom) this spinon configuration represents a multiplet with
$S_{T}=S_{i}^{\mathrm{orb}}=N_{s}/2$. Any distribution of spinons into multiple
orbitals represents more than one $S_{T}$-multiplet. The multiplets represented
are determined via the decomposition of tensor products of orbital spins.

For example, if we have one spinon with $m_{i}=2$ and three spinons with
$m_{i}=4$, that decomposition reads $\frac{1}{2}\otimes\frac{3}{2} =2\oplus1$.
The associated multiplets with $S_{T}=2$ and $S_{T}=1$ are found in motifs
three and thirteen, respectively. The configuration with two spinons in each of
the same two orbital represents three $S_{T}$-multiplets according to
$1\otimes1 =2\oplus1\oplus0$. They are found in motifs four, fourteen, and
twenty, respectively. The systematic application of these rules assigns a
unique spinon momentum and spin content to every $S_{T}$-multiplet.

The energetic split of different $S_{T}$-multiplets associated with the same
spinon momenta is caused by a coupling between orbital spins. In the
Haldane-Shastry model, which has higher symmetry, that coupling is absent and
these particular $S_{T}$-multiplets remain degenerate \cite{Hald88,Shas88}. In
the $XX$ model, which has lower symmetry, even the $S_{T}$-multiplets split up
energetically as will be discussed in Sec.~\ref{sec:XX} \cite{KMW08}.

The translation and reflection symmetries of $\mathcal{H}_{XXZ}$ dictate that
every state with wave number $k$ can be transformed into a state with wave
number $2\pi-k$ and the same energy, implying that any state with $k=0,\pi$ is
its own image. These transformation properties are reflected in the string
motif $(I_{\alpha}^{m}\to -I_{\alpha}^{m})$ and in the spinon motif $(m_{i}\to
N-m_{i})$ as is evident in Fig.~\ref{fig:stringN6}.

\subsection{Ising limit: stretched strings, domains, and 
solitons}\label{sec:Ising}
In the following we describe three ways of generating the spectrum of the Ising
chain, each producing a distinct set of quasiparticles with different exclusion
statistics. We begin with the analysis of the string solutions of the BAE from
Sec.~\ref{sec:axial}. To arrive at a non-divergent spectrum in the Ising limit
of $\mathcal{H}_{XXZ}$ we rescale the exchange coupling as follows:
\begin{equation}
  \label{eq:30plj3}
  \mathcal{H}_{I}\doteq\lim_{\Delta\to\infty}\Delta^{-1}\mathcal{H}_{XXZ} =
  J\sum_{l=1}^NS_l^zS_{l+1}^z. 
\end{equation}
The rescaled energy expression (\ref{eq:17plj3}) of the Bethe ansatz in raw
form thus becomes
\begin{equation}
  \label{eq:31plj3}
  \frac{E-E_{F}}{J}= -r
+\lim_{\Delta\to\infty}\sum_{i=1}^r\frac{\cos k_i}{\Delta}.
\end{equation}
Evidently, real magnon momenta contribute only summarily to the energy,
namely via the first term in (\ref{eq:31plj3}). Non-vanishing 
terms $\cos k_i/ \Delta$ can only come from complex $k_i$ with
infinite imaginary parts.

We first track the highest-weight states of the $S_{T}$-multiplets from the
$XXX$ limit to the Ising limit. For these states, all solutions that are real
at $\Delta=1$ stay real and all solutions that are complex at $\Delta=1$ have
imaginary parts that diverge as $\Delta\to\infty$. The non-highest-weight
states evolve far less uniformly. The additional rapidities, which are all
equal to $\pm\infty$ at $\Delta=1$, become, in general, finite and thus
contribute to the energetic split of the $S_{T}$-multiplets. In some states,
the extra $z_{i}$ stay real, in other states they combine to form complex
pairs. It appears that this transformation from real to complex rapidities
takes place throughout the axial regime. Here we focus on the end product at
$\Delta=\infty$, where the ferromagnetic domains in Ising product eigenstates
become the natural quasiparticles.

Expanding the trigonometric BAE (\ref{eq:1}) of the axial regime about the
Ising limit produces, in leading order, a set of linear BAE for the real parts
of the rapidities,
\begin{align}
  \label{eq:87}
  \left(N-\sum_{m'\in\mathcal{C}}t_{mm'}n_{m'}\right)z_{\alpha}^{m}
  = 2\pi J_{\alpha}^{m}-\sum_{m'\in\mathcal{C}}t_{mm'}\sum_{\alpha'=1}^{n_{m'}}
   z_{\alpha'}^{m'},
\end{align}
where the new set $\{J_{\alpha}^{m}\}$ of BQN, related to the original set
$\{I_{\alpha}^{m}\}$ by a shift that depends on $N$ and $n_{m}$, has the same
range (\ref{eq:10})
\begin{equation}
  \label{eq:31}
   |J_{\alpha}^{m}|\leq \frac{1}{2}\left(N-1-\sum_{m'\in\mathcal{C}}t_{mm'}
n_{m'}\right),\qquad t_{mm'}\doteq 2\mathrm{min}(m,m') -\delta_{mm'},
\end{equation}
which guarantees the correct number of states from the highest-weight type.
As the first step in the solution of the linear BAE (\ref{eq:87}) for these
states we introduce the quantities
\begin{align}
  \label{eq:88}
  \zeta_{m} \doteq N- \sum_{m'\in\mathcal{C}}n_{m'}t_{mm'},
  \qquad
  J_{m} \doteq \sum_{\alpha=1}^{n_{m}}J_{\alpha}^{m},
  \qquad
  Z_{m} \doteq \sum_{\alpha=1}^{n_{m}} z_{\alpha}^{m}.
\end{align}
Equations (\ref{eq:87}) summed over $\alpha$ can thus be brought into the form
\begin{align}
  \label{eq:90}
   Z_{m} = 2\pi\frac{J_{m}}{\zeta_{m}} - 
\sum_{m'\in\mathcal{C}}\frac{n_{m}}{\zeta_{m}} t_{mm'} Z_{m'}
\end{align}
with $\zeta_{m}$ guaranteed to be positive.
Equation (\ref{eq:90}) is solved by matrix inversion. Substitution of the
solution $Z_{m}$ into (\ref{eq:87}) yields the rapidities
\begin{align}
  \label{eq:91}
  z_{\alpha}^{m}
  = 2\pi \frac{J_{\alpha}^{m}}{\zeta_{m}}
  -\sum_{m'\in\mathcal{C}}\frac{t_{mm'}}{\zeta_{m}}Z_{m'}.
\end{align}
More explicit solutions in compact form for the situations where only 1-strings
or only 2-strings are present read
\begin{equation}
  \label{eq:35}
  z_{\alpha}^{1} = 
  \frac{2\pi}{N-r}J_{\alpha}^{1} 
  - \frac{2\pi}{N(N-r)}J_{1},
  \qquad \alpha=1,...,r,
\end{equation}
\begin{equation}
  \label{eq:36}
  z_{\alpha}^{2} = \frac{2\pi}{N-3r/2}J_{\alpha}^{2} -
  \frac{6\pi}{N(N-3r/2)}J_{2}, \qquad \alpha=1,...,r/2.
\end{equation}
If a single $m$-string with $1\leq m\leq N/2$ is present we have
$z_{1}^{m}=(2\pi/N)J_{1}^{m}$ with $|J_{1}^{m}|\leq (N-2m)/2$.

\begin{figure}[ht!]
\centerline{\includegraphics[width=13cm,angle=0]{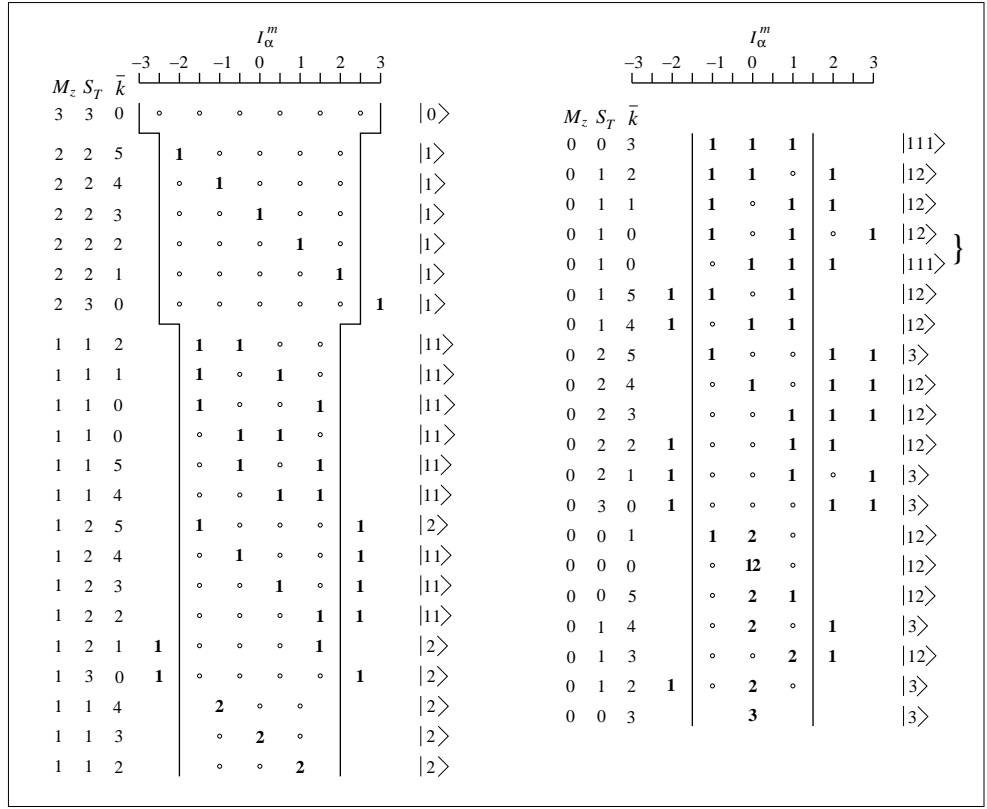}}
\caption{Specification of all $XXX$ eigenstates with $M_{z}\geq0$ for $N=6$ in
  the string representation (two columns). The quantum numbers $M_{z}$, $S_{T}$
  and $\bar{k}\doteq Nk/2\pi$ of each state are stated on the left. The
  vertical lines indicate the range (\ref{eq:10}) of the $I_{\alpha}^{1}$. The
  positions of the numbers $m=1,2,3$ between the vertical lines represent the
  values of the $I_{\alpha}^{m}$ for the highest-weight states (with
  $M_{z}=S_{T}$). The additional BQN of the non-highest-weight states (with
  $0\leq M_{z}<S_{T}$) are all located on the outside. In the Ising limit each
  state evolves into a translationally invariant linear combination of Ising
  product states with domains of lengths as indicated on the far right in each
  column.}
\label{fig:stringN6plus}
\end{figure}

In Fig.~\ref{fig:stringN6plus} we have reproduced the BQN of all 20
highest-weight states for $N=6$ at $\Delta=1$ from Fig.~\ref{fig:stringN6} and
have added the BQN for all 22 non-highest-weight states with $M_{z}\geq0$. The
additional $I_{\alpha}^{1}$ of non-highest-weight states, obtained for
$\Delta=1$ as described at the end of Sec.~\ref{sec:isotropic}, are located
beyond the range of the $I_{\alpha}^{1}$ for the highest-weight states
(vertical lines).

The choice of the extra $I_{\alpha}^{1}$ and the associated shift of the
$I_{\alpha}^{m}$ already present is not unique in most cases. Our choice was
guided by the aim to avoid multiple $I_{\alpha}^{1}$ of the same value and to
restore the symmetry relations described at the end of Sec.~\ref{sec:XXX}. For
states with $k=0,\pi$, which are their own images under the symmetry
transformation, neither goal was fully achievable for $N=6$. The
$I_{\alpha}^{1}$ configurations are asymmetric in several instances and two
identical $I_{\alpha}^{1}$ are unavoidable in one instance. These aesthetic flaws
pose no problem for what follows.

To the right of each string motif in Fig.~\ref{fig:stringN6plus} we have added
another motif in the shape of a ket. It identifies each string solution as
tracked from $\Delta=1$ to $\Delta=\infty$ with an Ising eigenstate as composed
of linear combinations of ferromagnetic domains. We have already introduced
these domains in the context of Fig.~\ref{fig:qpis22ar}. Here we use a notation
conducive to linking Ising product eigenstates with obvious domain composition
to Bethe eigenstates with specific string structure in the limit
$\Delta=\infty$.

Let the set of $\lambda$ product vectors with $r=N/2-M_{z}$ flipped spins that are
generated from one such state, $|\sigma_{1}\cdots\sigma_{N}\rangle$, via translations be
represented by $\{|\sigma_{1}\cdots\sigma_{N} \rangle\}_{\lambda}$.
Let 
\begin{equation}
  \label{eq:2}
  \big\{\big|\underbrace{1\cdots1}_{\nu_{1}}
\underbrace{2\cdots2}_{\nu_{2}}\underbrace{3\cdots3}_{\nu_{3}}
\cdots\big\rangle\big\}_{\Lambda}
\end{equation}
represent the set of $\Lambda$ translationally invariant linear combinations of
all product states that contain $\nu_{n}$ domains of length $n$ subject to the
constraint $\nu_{1}+2\nu_{2}+3\nu_{3}+\cdots =r$. Any set of the latter kind is
constructed from one or several sets of the former kind with matching domain
content. For $N=6$ there are seven sets of translationally invariant Ising
eigenstates at $M_{z}\geq0$ thus constructed:
\begin{subequations}\label{eq:91sub}
\begin{align}
 & \{|0\rangle\}_{1}  \stackrel{l.c.}{\leftarrow}  
\{|\uparrow\uparrow\uparrow\uparrow\uparrow\uparrow\rangle\}_1;\qquad 
\{|1\rangle\}_{6} \stackrel{l.c.}{\leftarrow}
\{|\uparrow\uparrow\uparrow\uparrow\uparrow\downarrow\rangle\}_6;\qquad 
\{|2\rangle\}_{6} \stackrel{l.c.}{\leftarrow}
\{|\uparrow\uparrow\uparrow\uparrow\downarrow\downarrow\rangle\}_6; 
\\
& \{|11\rangle\}_{9}  \stackrel{l.c.}{\leftarrow}
\{|\uparrow\uparrow\uparrow\downarrow\uparrow\downarrow\rangle\}_6,
\{|\uparrow\uparrow\downarrow\uparrow\uparrow\downarrow\rangle\}_3;\qquad
\{|3\rangle\}_{6} \stackrel{l.c.}{\leftarrow}
\{|\uparrow\uparrow\uparrow\downarrow\downarrow\downarrow\rangle\}_6;
\\
& \{|12\rangle\}_{12}  \stackrel{l.c.}{\leftarrow}
\{|\uparrow\uparrow\downarrow\downarrow\uparrow\downarrow\rangle\}_6,
\{|\uparrow\uparrow\downarrow\uparrow\downarrow\downarrow\rangle\}_6;\qquad 
\{|111\rangle\}_{2} \stackrel{l.c.}{\leftarrow}
\{|\uparrow\downarrow\uparrow\downarrow\uparrow\downarrow\rangle\}_2. 
\end{align}
\end{subequations}

In Fig.~\ref{fig:stringN6plus} we have identified each of these 42 states with
a particular BAE solution in the limit $\Delta\to\infty$. The identification is
straightforward for the highest-weight states. Each BQN $I_{\alpha}^{m}$
represents exactly one domain of size $\mu=m$. However, the evolution of the
non-highest-weight states between $\Delta=1$ and $\Delta\to\infty$ is far less
predictable. The extra BQN are all of the type $I_{\alpha}^{1}$ at $\Delta=1$
and the associated rapidities are real and infinite. In some cases all extra
rapidities stay real.  The domain structure of these states is determined by
the $I_{\alpha}^{m}$ just as in highest-weight states.

However, for the majority of non-highest-weight states some of the rapidities
that start out real at $\Delta=1$ turn into complex-conjugate pairs at
$\Delta>1$ with imaginary parts that diverge as $\Delta\to\infty$. This
metamorphosis necessitates a reconfiguration of the associated BQN (not shown
in Fig.~\ref{fig:stringN6plus}). Of the 22 non-highest-weight states for $N=6$
only six do not acquire additional complex rapidities between $\Delta=1$ and
$\Delta\to\infty$, namely those states for which the string content encoded in
the $I_{\alpha}^{m}$ matches the domain content encoded in the ket.

The identification of the BAE solutions in the limit $\Delta\to\infty$ with
Ising eigenstates of specific domain content may be less certain in longer
chains where states with different $\{\nu_{\mu}\}$ but equal $M_{z}$ are
degenerate. That is the case for the two sets of states $\{|13\rangle\}_{24}$
and $\{|22\rangle\}_{12}$ in a chain with $N=8$ sites, for example. However, it
seems reasonable to assume that this degeneracy is absent at $\Delta<\infty$
in most if not all cases and thus guarantees that the BAE solutions have, in
general, a unique domain content in the limit $\Delta\to\infty$.  Even though
the systematics of the transformation of some BAE solutions between $\Delta=1$
and $\Delta\to\infty$ appears elusive at present, the systematics of the
endproduct, namely the domain structure of the Ising spectrum for arbitrary
$N$, is well known.

An Ising chain of length $N$ can accommodate domains with $\mu=1,\ldots,N-1$
consecutive flipped spins. Domains of size $\mu$ are treated as distinct
species of independent particles. The capacity for domains is restricted by
the relation
\begin{equation}
  \label{eq:55clj}
  \sum_{\mu=1}^{N-1} (\mu+1) \nu_\mu\leq N.
\end{equation}
The total number of Ising eigenstates with domain content
$\{\nu_{1},\nu_{2},\ldots\}$ is governed by a multiplicity expression somewhat
similar to (\ref{eq:19}) yet different \cite{LVP+08}:
\begin{equation}
  \label{eq:57clj}
   W(\{\nu_\mu\}) = \frac{N}{N-r}\;\prod_{\mu=1}^{N-1}   
\left(
\begin{tabular}{c}
$d_\mu +\nu_\mu-1$ \\ $\nu_\mu$
\end{tabular}
\right),\qquad 
d_\mu=A_{\mu} -\sum_{\mu'=1}^{N-1} g_{\mu\mu'}(\nu_{\mu'}-\delta_{\mu\mu'}),
\end{equation}
\begin{equation}
  \label{eq:59clj}
  A_{\mu}=N-\mu,\qquad g_{\mu\mu'}=\left\{
\begin{tabular}{ll}
$\mu'$, & $\mu<\mu'$, \\ $\mu'+1$, & $\mu\geq\mu'$
\end{tabular}
\right.,\qquad  r= \sum_{\mu=1}^{N-1} \mu \nu_{\mu}.
\end{equation}
The capacity of the system for domains is controlled by $A_{\mu}$ and the
statistical interaction between domains by $g_{\mu\mu'}$.  All domains have the
same energy, $J$. The statistical mechanics of domains for the Ising chain has been
carried out exactly \cite{LVP+08}, reproducing familiar results.

In the context of Fig.~\ref{fig:qpis22ar} we have already qualitatively
described the antiferromagnetic domain walls (solitons) that are complementary
to the ferromagnetic domains.  Among the four distinct bonds in the general
product state $|\sigma_1\sigma_2\cdots\sigma_N\rangle$, the bonds
$\uparrow\uparrow$, $\downarrow\downarrow$ represent solitons with spin $+1/2$,
$-1/2$, respectively, and $\uparrow\downarrow$, $\downarrow\uparrow$ are vacuum
bonds.  Close-packed solitons with like spin orientation reside on successive
bonds (e.g.  $\uparrow\uparrow\uparrow$), whereas close-packed solitons with
opposite spin orientation are separated by one vacuum bond (e.g.
$\uparrow\uparrow\downarrow\downarrow$). Each of the seven sets of Ising
eigenstates (\ref{eq:91sub}) contains a specific number of spin-up and
spin-down solitons, $(N_{+},N_{-})= (6,0), (4,0), (3,1), (2,0), (2,2), (1,1),
(0,0).$ All solitons have the same energy, $J/2$.  The multiplicity expression
for solitons \cite{LVP+08},
\begin{equation}
  \label{eq:27}
   W(N_+,N_-) = \frac{2N}{N-N_a}\;\prod_{\sigma=\pm}  
\left(
\begin{tabular}{c}
$d_\sigma +N_\sigma -1$ \\ $N_\sigma$
\end{tabular}
\right),\qquad d_\sigma=A_{\sigma} 
-\sum_{\sigma'=\pm}g_{\sigma\sigma'}(N_{\sigma'}-\delta_{\sigma\sigma'}),
\end{equation}
\begin{equation}
  \label{eq:32}
  A_{\sigma}=\frac{1}{2}(N-1),\qquad g_{\sigma\sigma'}=\frac{1}{2},\qquad
  N_{a}=N_{+}+N_{-}, 
\end{equation}
is similar to expression (\ref{eq:25}) for spinons. Solitons and spinons are
both semions but have different pseudovacua. Whereas the spinon vacuum was
found to be unique, the soliton vacuum is twofold, consisting of the two
product N{\'e}el states, $|\uparrow\downarrow\uparrow\cdots\downarrow\rangle$
and $|\downarrow\uparrow\downarrow\cdots\uparrow\rangle$ or linear combinations
thereof. The soliton vacuum, like the spinon vacuum, is realized only in chains
with even $N$.

\subsection{XX limit: broken strings, fermions, and 
spinons}\label{sec:XX}

Here we investigate what happens to the highest-weight and non-highest-weight
states of the $S_{T}$-multiplets in the
presence of planar exchange anisotropy, particularly in the limit $\Delta=0$.
We have seen that in the axial regime there exists a tendency for real BAE
solutions to become complex. As $\Delta\to\infty$ all imaginary parts diverge,
binding the magnons tightly into domains.
In the planar regime, there exists a trend in the opposite direction.  At
$\Delta=0$ all BAE solutions can be regularized as stated in
Sec.~\ref{sec:planar}, making all magnon momenta real. All strings with $m>1$
break up into 1-strings. The configurations allowed by the set (\ref{eq:34}) of
BQN, which describe regularized solutions, produce the complete spectrum for
arbitrary $N$.

\begin{figure}[ht!]
\centerline{\includegraphics[width=13cm,angle=0]{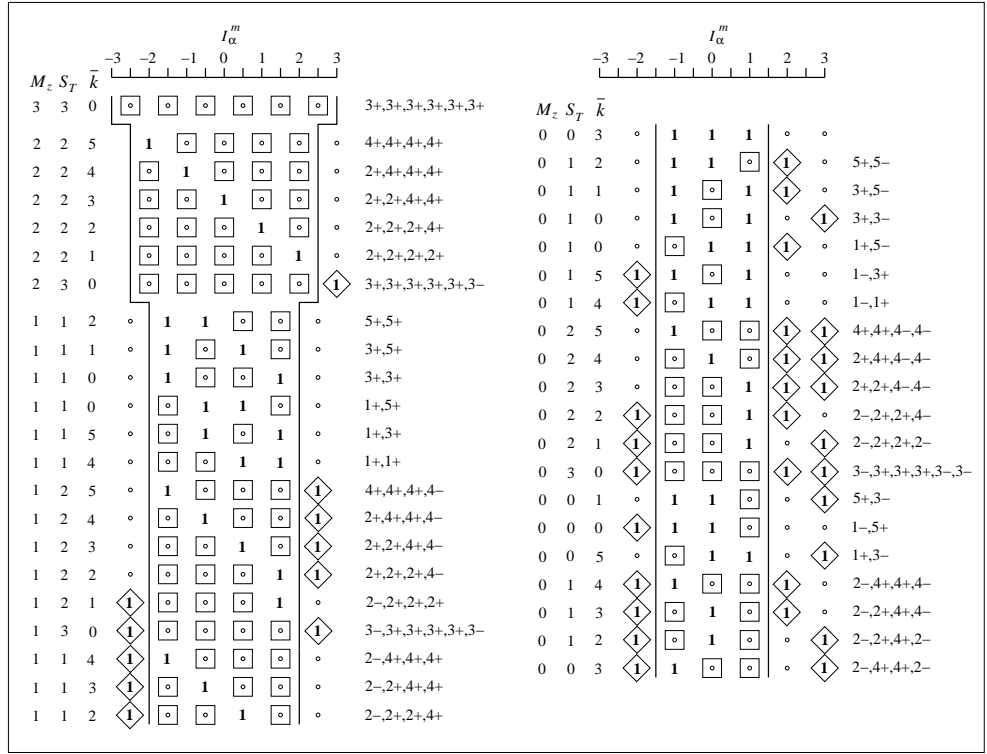}}
\caption{Specification of all $XX$ eigenstates with $M_{z}\geq0$ for $N=6$ in
  the string representation (two columns). The quantum numbers $M_{z}$, and
  $\bar{k}\doteq Nk/2\pi$ of each state are stated on the left. Also shown on
  the left is the quantum number $S_{T}$ of the multiplet at $\Delta=1$ from
  which the state evolved. The states are listed in the same sequence as those
  in Fig.~\ref{fig:stringN6plus}. The positions of the '1' represent the values
  of the $I_{\alpha}^{1}$ from (\ref{eq:34}) and constitute the motifs of the
  (broken) string configuration of the $XX$ eigenstates. The motif of the
  complementary spinon configuration is marked by the squares (spin-up spinons)
  and diamonds (spin-down spinons) as explained in the text.The momentum
  quantum number $m_{i}$ and the spin orientation $\sigma_{i}$ of each spinon
  present in that state are given to the right of the dual string/spinon motif.
}
\label{fig:stringN6minus}
\end{figure}

In Fig.~\ref{fig:stringN6minus} we show the configurations of $I_{\alpha}^{1}$
of all 42 eigenstates with $M_{z}\geq0$ at $\Delta=0$ for $N=6$ in the same
sequence as the corresponding $S_{T}$-multiplet states at $\Delta=1$ have been
listed in Fig.~\ref{fig:stringN6plus}. All multiplet states at $\Delta=1$ that
contain only 1-strings keep the same configuration at $\Delta=0$. In the other
states, all $I_{\alpha}^{1}$ stay in the same position except the pair of
$I_{\alpha}^{1}$ that have identical values in Fig.  \ref{fig:stringN6plus}.
They are replaced in Fig.~\ref{fig:stringN6minus} by a pair of distinct
$I_{\alpha}^{1}$.  Furthermore, each $I_{\alpha}^{2}$ in
Fig.~\ref{fig:stringN6plus} is replaced by two $I_{\alpha}^{1}$ in
Fig.~\ref{fig:stringN6minus} and the one occurrence of a $I_{\alpha}^{3}$ is
replaced by three $I_{\alpha}^{1}$.  Whereas the general rules for these
substitutions still elude us, the constraints imposed by symmetry and
conservation laws eliminate any ambiguity for the case $N=6$ shown here. The
resulting configuration of $I_{\alpha}^{1}$ produces, via (\ref{eq:33}), the
exact configurations of Jordan-Wigner fermion momenta in a periodic chain
\cite{LSM61,MBA71}.  The statistical mechanics of the regularized 1-strings is
that of a system of free Jordan-Wigner fermions \cite{Kats62}.

Each $XX$ eigenstate thus identified by its composition of 1-strings has a
unique composition of spinons. The rules for inferring the spinon motif from
the string motif are consistent with the rules described in
Sec.~\ref{sec:isotropic} for the case $\Delta=1$ but there are some noteworthy
differences. The lower rotational symmetry at $\Delta=0$, which splits up the
$S_{T}$-multiplet degeneracy, makes it possible to assign to each $XX$ state
not only a spinon configuration with unique momentum quantum numbers but to the
spinons in each orbital also a unique spin orientation.

The spinon motif is encoded in the string motif of each state shown in
Fig.~\ref{fig:stringN6minus} as described in the following. (i) Consider the
vertical lines dividing the space of the $I_{\alpha}^{1}$ into two domains the
inside and the outside, the latter wrapping around at $\pm N/2$. (ii) Every
$I_{\alpha}^{1}$-vacancy (small circle) inside represents a spin-up spinon
(marked by a square) and every $I_{\alpha}^{1}$ ('1') outside represents a
spin-down spinon (marked by a diamond). (iii) Any number of adjacent spinons in
the motif are in the same orbital, i.e. have the same momentum quantum number
$m_{i}$ from the set (\ref{eq:23}).  Two squares or diamonds that are separated
by $\ell$ consecutive '1's have spinon quantum numbers separated by $2\ell$.
(iv) The spinon momentum quantum numbers are sorted in increasing order from
the line on the left toward the right through the inside domain $(m_{i}^{+})$
and toward the left with wrap-around through the outside domain $(m_{i}^{-})$.

The spinon motif of Fig.~\ref{fig:stringN6minus} establishes the much needed
link between spinon motif of the fermion representation at $\Delta=0$ introduced in
Refs.~\cite{AKMW06,KMW08} and the spinon motif of the string representation at
$\Delta=1$ introduced in Fig.~\ref{fig:stringN6plus}. Unlike the solitons at
$\Delta\to\infty$, the spinons at $\Delta=0$ are not free. Nevertheless, an exact
statistical mechanical analysis of spinons at $\Delta=0$ is possible
\cite{KMW08}. Away from the Ising limit, the solitons become interacting
particles as well. They can be tracked all the way from $\Delta\to\infty$ to
$\Delta=0$. A comparison between the spinon composition and soliton composition
of the $XX$ eigenstates can be found in Ref.~\cite{LVP+08}.

\section{Conclusion}\label{sec:concl}
The long-established integrability of $\mathcal{H}_{XXZ}$ imposes stringent
constraints on the nature of the interaction between quasiparticles of any
kind, reducing it, effectively, to two-body scattering events. As a
consequence, these quasiparticles have infinite lifetimes and can thus be
regarded as structural elements of the $XXZ$ eigenstates. This makes it
possible to systematically generate the complete spectrum from the pseudovacuum
of one or the other set of quasiparticles.

In this work we have discussed two complementary sets of quasiparticles, one
set being the basic elements of string solutions of the coordinate Bethe ansatz
and the other the semionic spinons or solitons. Our focus has been on
developing interlocking motifs for each set of quasiparticles for the purpose
of tracking them across the axial and planar regimes of $\mathcal{H}_{XXZ}$.
Starting from the symmetry point $\Delta=1$ in parameter space we have
identified opposite trends, in the two regimes, in the evolution of string
particles, which can be thought of as bound clusters of magnons.  In the axial
regime, the strings show a tendency of increasing tightness in the binding
within each cluster (rapidities with growing imaginary parts) and a tendency of
merging clusters (real rapidities becoming complex). The endproduct at
$\Delta=\infty$ are ferromagnetic domains of flipped spins. In the planar
regime, the strings appear to loosen up and break apart. At $\Delta=0$ it is
possible to transform away all complex rapidities and with them any trace of
magnon clustering. The free magnons behave like hard-core bosons or free
fermions.

For $J>0$ the pseudovacuum of string particles is located at the top of the
spectrum throughout the axial regime and then migrates down toward the center
of the spectrum in the planar regime. The ground state (physical vacuum) of
$\mathcal{H}_{XXZ}$ is unique in the planar regime and twofold in the axial
regime. It can be identified as the pseudovacuum of spinons or solitons,
respectively. We have given detailed descriptions of spinons at $\Delta=0$ and
$\Delta=1$ and of solitons at $\Delta=\infty$, including their relationship to
the string particles.  What remains to be investigated is the metamorphosis of
the string particles between the special points $\Delta=0,1,\infty$ in
parameter space. This is work in progress. Of particular interest is an
analysis of the transformations for which we have given here only qualitative
descriptions and the impact of these transformations on the motifs such as
displayed in Figs.~\ref{fig:stringN6}-\ref{fig:stringN6minus} including the
effects on the complementary spinon and soliton particles.


\begin{thebibliography}{100}

\bibitem{SVW76}
Steiner M, Villain J and Windsor C~G, 1976 \emph{Adv. Phys.} \textbf{25} 87

\bibitem{Furr00}
Furrer A (Ed.), \emph{Frontiers in neutron scattering} (World Scientific,
  Singapore, 2000)

\bibitem{Beth31}
Bethe H, 1931 \emph{Z. Phys.} \textbf{71} 205

\bibitem{Batc07}
Batchelor M~T, 2007 \emph{Phys. Tod.} \textbf{60/1} 36

\bibitem{LL63}
Lieb E and Liniger W, 1963 \emph{Phys. Rev.} \textbf{130} 1605

\bibitem{YY69}
Yang C~N and Yang C~P, 1969 \emph{J. Math. Phys.} \textbf{10} 1115

\bibitem{Taka99}
Takahashi M, \emph{Thermodynamics of one-dimensional solvable models}
  (Cambridge University Press, 1999)

\bibitem{LW68}
Lieb E~H and Wu F~Y, 1968 \emph{Phys. Rev. Lett.} \textbf{20} 1445

\bibitem{EFG+05}
Essler F~H~L, Frahm H, G{\"o}hmann F, Kl{\"u}mper A et~al., \emph{The
  one-dimensional {H}ubbard model} (Cambridge University Press, 2005)

\bibitem{BT79}
Bergknoff H and Thacker H~B, 1979 \emph{Phys. Rev. Lett.} \textbf{42} 135

\bibitem{KM97}
Karbach M and M{\"u}ller G, 1997 \emph{Comp. in Phys.} \textbf{11} 36

\bibitem{BKMW04}
Biegel D, Karbach M, M\"uller G and Wiele K, (2004) \emph{Phys. Rev. B}
  \textbf{69} 174404

\bibitem{AKMW06}
Arikawa M, Karbach M, M{\"u}ller G and Wiele K, 2006 \emph{J. Phys. A: Math.
  Gen.} \textbf{39} 10623

\bibitem{KMW08}
Karbach M, M{\"u}ller G and Wiele K, 2008 \emph{J. Phys. A: Math. Gen.}
  \textbf{41} 205002

\bibitem{DFM01}
Deguchi T, Fabricius K and McCoy B~M, 2001 \emph{J. Stat. Phys.} \textbf{102}
  701

\bibitem{FM01}
Fabricius K and McCoy B~M, 2001 \emph{J. Stat. Phys.} \textbf{103} 647

\bibitem{FM01a}
Fabricius K and McCoy B~M, 2001 \emph{J. Stat. Phys.} \textbf{104} 573

\bibitem{Hald91a}
Haldane F~D~M, 1991 \emph{Phys. Rev. Lett.} \textbf{67} 937

\bibitem{LSM61}
Lieb E, Schulz T and Mattis D, 1961 \emph{Ann. Phys.} \textbf{16} 407

\bibitem{Kats62}
Katsura S, 1962 \emph{Phys. Rev.} \textbf{127} 1508

\bibitem{MBA71}
McCoy B~M, Barouch E and Abraham D~B, 1971 \emph{Phys. Rev. A} \textbf{4} 2331

\bibitem{Derz02}
Derzhko O, 2002 \emph{Cond. Mat. Phys.} \textbf{5} 729

\bibitem{CG66}
des Cloizeaux J and Gaudin M, 1966 \emph{J. Math. Phys.} \textbf{7} 1384

\bibitem{Wu94}
Wu Y~S, 1994 \emph{Phys. Rev. Lett.} \textbf{73} 922

\bibitem{FS98}
Frahm H and Stahlsmeier M, 1998 \emph{Phys. Lett. A} \textbf{250} 293

\bibitem{LVP+08}
Lu P, Vanasse J, Piecuch C, M{\"u}ller G et~al., 2008 \emph{J. Phys. A: Math.
  Gen.} \textbf{41} 265003

\bibitem{Hald88}
Haldane F~D~M, 1988 \emph{Phys. Rev. Lett.} \textbf{60} 635

\bibitem{Shas88}
Shastry B~S, 1988 \emph{J. Stat. Phys.} \textbf{50} 57

\end{thebibliography}

\label{last@page}
\end{document}